\documentclass[preprint,12pt]{elsarticle}

\usepackage{amssymb}

\usepackage{verbatim}
\usepackage{xcolor}
\usepackage{float}
\usepackage{hyperref}

\journal{arXiv}

\begin{document}

\begin{frontmatter}



\title{Achieving a Data-driven Risk Assessment Methodology for Ethical AI}


\author[inst1]{Anna Felländer}

\affiliation[inst1]{organization={AI Sustainability Center},
            city={Stockholm},
            country={Sweden}}

\author[inst1,inst2]{Jonathan Rebane}


\author[inst1,inst3]{Stefan Larsson}

\author[inst1,inst5]{Mattias Wiggberg}

\author[inst1,inst4]{Fredrik Heintz}

\affiliation[inst2]{organization={Stockholm University},
            addressline={Department of Computer and Systems Sciences}, 
            city={Stockholm},
            country={Sweden}}

\affiliation[inst3]{organization={Lund University},
            addressline={Department of Technology and Society}, 
            city={Lund},
            country={Sweden}}
            
\affiliation[inst4]{organization={Linköping University},
            addressline={Department of Computer and Information Science}, 
            city={Linköping},
            country={Sweden}}
            
\affiliation[inst5]{organization={KTH Royal Institute of Technology},
            addressline={Department of Industrial Engineering}, 
            city={Stockholm},
            country={Sweden}}

\begin{abstract}
The AI landscape demands a broad set of legal, ethical, and societal considerations to be accounted for in order to develop ethical AI (\textbf{eAI}) solutions which sustain human values and rights.
Currently, a variety of guidelines and a handful of niche tools exist to account for and tackle individual challenges. However, it is also well established that many organizations face practical challenges in navigating these considerations from a risk management perspective. Therefore, new methodologies are needed to provide a well-vetted and real-world applicable structure and path through the checks and balances needed for ethically assessing and guiding the development of AI. In this paper we show that a multidisciplinary research approach, spanning cross-sectional viewpoints, is the foundation of a pragmatic definition of ethical and societal risks faced by organizations using AI. Equally important is the findings of cross-structural governance for implementing eAI successfully. Based on evidence acquired from our multidisciplinary research investigation, we propose a novel data-driven risk assessment methodology, entitled DRESS-eAI. In addition, through the evaluation of our methodological implementation, we demonstrate its state-of-the-art relevance as a tool for sustaining human values in the data-driven AI era.
\end{abstract}




\end{frontmatter}


\section{Introduction}
\label{sec:intro}

The evolving data-driven technology sector has resulted in AI solutions becoming pervasively implemented throughout much of society. Such implementations demand a myriad of legal, ethical, and societal considerations which must be accounted for in order to develop ethical AI (\textbf{eAI}) solutions which sustain human values in an emerging data-driven era \cite{cath2018governing}. The cost of ignoring eAI issues can be very high, with several high-profile AI systems ultimately needing to be shut down after risks inadvertently materialized and massive reputational losses occurred \cite{lauer2021you}. 

As of today, most means of risk management in the eAI landscape consists of a variety of guidelines, recommendations, and a handful of niche tools to account for and tackle individual challenges \cite{hagendorff2020ethics, jobin2019global, bellamy2018ai, canca2020operationalizing, larsson2020governance}. However, such resources have been criticized for being too abstract or technology-centered, lacking a direct focus on organizational viewpoints and needs \cite{theodorou2020towards}, such as the compatibility with standardised risk assessment models. In addition, existing methodologies for eAI focus on the life-cycle development process rather than risk assessment \cite{d2018towards} or lack validation, emphasis on human rights, and cross-sector perspectives \cite{brendel2021ethical}. Even if there are proposals for self-assessment \cite{HLEG}, and structured approaches on "ethics-based auditing" \cite{brown2021algorithm, floridi2019unified}
, requests have been made for the development and validation of methods that can be applied in reality to provide guidance on assessing and managing eAI organisation risk. \cite{wright2020ai, theodorou2020towards, brendel2021ethical} 

From an organizational viewpoint, risk assessment methodologies for technical systems exist to provide a linear structure for identifying and mitigating unregulated business risks with individualized risk assessment phases \cite{pandey2012comparative}. However, what is needed are novel approaches, developed from a holistic multidisciplinary approach which incorporates technical, legal and societal perspectives, with the objective of detecting negative eAI externalities of organizations that would otherwise infringe on legal and human rights alongside organizational principles. Research gaps needed to be filled are in relation to multi-stakeholder organizational perspectives for establishing core concepts with regard to the eAI risk landscape \cite{rodrigues2020legal}. Such knowledge can then be leveraged as a basis for filling a gap in relation to the development of an eAI risk assessment methodology. Due to the rapidly progressing eAI landscape, such a methodology must be flexible in the sense that defined risks, concepts and methods are flexible enough to accommodate an evidence-based evolution.

It is well-established that many organizations face practical challenges in navigating eAI considerations \cite{lauer2021you, rakova2021responsible, desouza2020designing}. There is a general discussion regarding how ethical AI in organisations could be handled \cite{clarke2019world}. Yet the presence of proof-of-concepts where real organisations and real data have been tested is low. Therefore, new methodologies are needed to provide a  structure and path through the checks and balances needed for ethically assessing an AI. The question we wish to answer in this paper is: \textit{\textbf{How can a standardized approach to ethical AI risk assessment be constructed that is compatible with organizational demands across a large variety of contexts?}}

Our contributions which aim to address this question are as follows: 

\begin{itemize}

\item Reporting the findings of a multidiciplinary perspective research investigation. This investigation was initiated as a eAI landscape review of risks. Then followed by cross-sector expert discussions which provided categorizations of unintended root causes of risks, which we call pitfalls. These discussions also identified fundamentals, which must be enacted for organizations within the eAI domain, in order to prevent pitfalls. These results helped to formulate the requirements needed for a standardized eAI risk assessment methodology, that is compatible with regulatory demands across a large variety of contexts. 
\item A data-driven methodology as a means to ensure that human values and rights are sustained for data-driven AI applications based on these results.

\item Leveraging these discovered results to present a data-driven methodology implemented in the real world as a means to help ensure human values and rights are sustained in the data-driven AI era.

\item An initial evaluation of the methodology implementation in the context of two case studies that shows the effectiveness of the methodology along with its data-driven development through iterative improvements.

\end{itemize}

\section{Research Approach}
\label{sec:research}

To answer the question of how to provide an approach to eAI risk assessment that is compatible with different organization demands, we systematically approached the problem from a multi-disciplinary research perspective, which is further clarified below, to establish requirements needed for a well-vetted and real-world applicable risk assessment methodology in eAI. We specify the importance of a multi-disciplinary approach as vital in order to capture computer scientific notions of AI as well as humanistic and  social-scientific notions of ethics-based governance. This process was based on steps which has been visualized in Figure 1 and specified in detail below. 

\begin{figure}[ht!]
    \centering
    \includegraphics[width=1\columnwidth, height=3.2in]{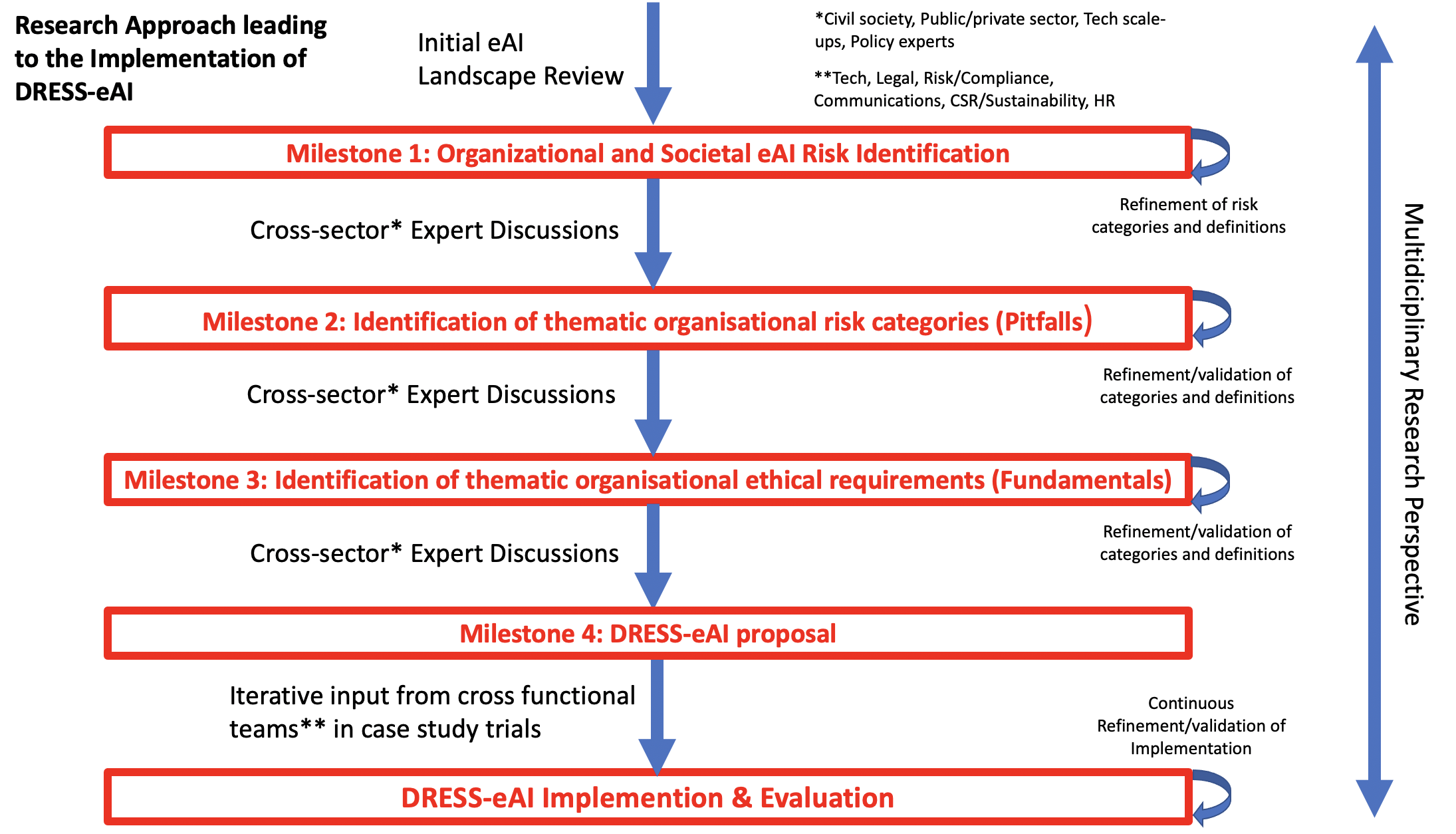}
    \vspace{-2mm}
    \caption{Multidisciplinary Research Approach leading to the development, implementation and evaluation of the proposed DRESS-eAI Methodology 
    \label{fig:research}}
\end{figure}

\subsection{Organizational and Societal eAI Risk Identification}

 To establish the needs of a risk assessment methodology for organizational eAI risks which pose a threat to human values and rights, a systemic research process was conducted. As a first step, a literature review was performed \citep{larsson2019sustainable}. This review included both a quantitative and qualitative assessment of the literature on fairness, transparency and accountability in AI, in order to exhaustively examine the eAI landscape while identifying and discussing societal eAI risks which do not cause intentional harm . The review helped to form a springboard for future expert exercises on the topic.  One outcome of this study was a realisation that much of the eAI landscape places a strong focus on technical rather than organisational risks. This realisation motivated the need for highlighting organisational risks with society as the primary stakeholder.

Through an initial evaluation and iterative expert reviews of the topic, a list of the most common and distinctive risks related to eAI was composed. This consisted of eight organizational risks, whose definitions are backed by global human rights legislation and other external analysis \cite{assembly1948universal, meek2016managing}, including the commonality of issues addressed in ethics guidelines \cite{jobin2019global}:

\begin{itemize}
    \item \textbf{Privacy intrusion} -- AI and data driven solutions interfering with personal or sensitive data without regarding:  consent of the individual or groups whose data is collected, how data is shared or stored, agreement of the law, or other legitimate needs to protect the best interests of an individual or groups. (Right to privacy) 
    \item \textbf{Amplified discrimination} -- AI and data driven solutions which cause, facilitate, maintain, or increase prejudicial decisions or treatment and/or biases towards race, sex, or any other protected groups obliged to equal treatment. (Right to fair treatment)      
    \item \textbf{Violation of autonomy and independent decision making} -- AI and data driven solutions which intentionally or unintentionally, and without consent, facilitate behavioural changes that manipulate independent decision making and social well-being. (Right to autonomy)
    \item \textbf{Social exclusion and segregation} -- AI and data driven solutions contributing to or maintaining an unfair denial of: resources, rights, goods,  and ability to participate in normal relationships and activities, whether in economic, social, cultural, organisational or political arenas. (Right to inclusion)
    \item \textbf{Harm to Safety} -- AI and data driven solutions facilitating unwanted physical harms to an individual or organization stemming from underdeveloped AI, and attributed to negligence from an organisation.  (Right to physical safety)
    \item \textbf{Harm to Security of Information} -- AI and data driven solutions facilitating potential damage from unauthorized access of private data, due to faulty data protection and processing, or criminal activity.  (Right to security of information) 
    \item \textbf{Misinformation and Disinformation} -- AI and data driven solutions which intentionally or unintentionally distribute information that has universally been declared as false and harmful to society. (Right to be informed)
    \item \textbf{Prevention of access to public service} -- AI and data driven solutions contributing to or maintaining a denial of public social assistance and service (Right to public service access)
\end{itemize}

\subsection{Identification of thematic organizational risk source categories (Pitfalls)}

The output of the initial eAI landscape review formed the basis for a series of cross-sector expert-based exercises in which thematic categories of root causes of eAI risks and in an organizational context were identified.  The aim of the exercises were to answer the following questions which arose from the landscape review:

\begin{itemize}

    \item How should we define thematic categories as root causes of eAI risks in an organizational and societal context?

    \item How can these risks be mitigated from broad society and organizational based perspectives?

\end{itemize}

These exercises were conducted as part of an initiative together with the Swedish Innovation Agency (Vinnova), a government agency that administers state funding for research and development. Starting in 2018, participants were gathered to perform a series of exercises in Stockholm, Sweden which involved cross-sector experts in the domains of civil society, public/private sector, tech scale-ups and policy. More specifically, this included Vinnova-linked organizations spanning legal, technical, business, communication, and sustainability/CSR from public and private domains, along with organisations associated with the AI Sustainability Center, a start-up based in Stockholm, Sweden. A full list of these organisations can be seen in Table \ref{table:participatingExpertPanelists}. Exercises were performed with the listed organisations, over the course of several years, consisting of round-table discussions, panel discussions, seminars, and joint analyses of ethical AI topics. Evidence continues to be collected through use-cases to ensure indentified categories are exhaustive of the eAI risk landscape from societal and organisation perspectives.




\begin{table}[ht!]

\begin{tabular}{ |p{10cm}|p{3cm}|  }
 \hline
 \textbf{Organisation}  & \textbf{Sector} \\
 \hline
 Swedish institute for Standards  & Cross-Sector\\
 \hline
 City of Stockholm & Public\\
 \hline
 City of Malmö & Public\\
 \hline
 Sana Labs (AI for individualised learning) & Tech start-up\\
 \hline
 Swedish Tax Agency & Public\\
 \hline
 Cirio (Law firm)  & Legal\\
 \hline
 Microsoft Sweden & Private Tech\\
 \hline
 Google Sweden & Private Tech\\
 \hline
 Boston Consulting Group & Cross-sector\\
 \hline
 Telia & Private Tech\\
 \hline
 Human Rights Watch & Human Rights\\
 \hline
 Ericsson & Private Tech\\
 \hline
The Institute for Futures Studies  & Public Research\\
 \hline
Karolinska Institute & Public Research\\
 \hline
Royal Institute of Technology (KTH) & Public Research\\
 \hline
Stockholm School of Economics & Public Research\\
 \hline
Civil Rights Defenders & Human Rights\\
 \hline
 Södertörn University & Public Research\\
 \hline
\end{tabular}
\caption{Participating organisations in the cross-sector expert exercises}
\label{table:participatingExpertPanelists}
\end{table}

This selection was based on a need to form a multidisciplinary organizational perspective on societal, ethical, and legal considerations towards the eAI risk landscape. The goal of exercises performed was to reflect on how the identified eAI risks would appear in each of these domains and identify which thematic categories form the root cause of each of these risks across all domains. Through this process, the experts reached a consensus on 4 common themes, which we refer to as \textbf{pitfalls}. These 4 pitfalls are:

\begin{itemize}
    \item \textbf{Misuse/overuse of Data} -- The AI application/solution could be overly intrusive, using private data, or it could be used for unintended purposes by others.
    \footnote{Further discussion on Misuse/overuse of Data: \cite{brundage2018malicious, larsson2021ai}}
    \item \textbf{Bias of the Creator} -- Values and bias are intentionally or unintentionally programmed by the creator who may also lack knowledge/skills of how the solution could scale in a broader context.
    \footnote{Further discussion on Bias of the Creator: \cite{whittaker2019disability, noble2018algorithms}}
    \item \textbf{Immature Data \& AI} -- Insufficient training of algorithms on data sets as well as lack of representative data. could lead to incorrect and unethical recommendations. \footnote{Immature Data \& AI examples: \cite{buolamwini2018gender, shankar2017no, larsson2019socio}}
    \item \textbf{Data bias} -- The data available is not an accurate reflection of reality or the preferred reality and may lead to incorrect and unethical recommendations. \footnote{Data bias examples: \cite{buolamwini2018gender, shankar2017no} }
\end{itemize}

\subsection{Identification of thematic organizational ethical requirements (Fundamentals)}

The next step of in the series of excerises was to discuss thematic categories for how to prevent and overcome such pitfalls in an organizational context. The results of this were the establishment of organization structural eAI foundations as thematic categories. Through this process, the experts identified 4 common themes, which we refer to as \textbf{fundamentals}, and is also echoed in much recent principled work on AI \cite{jobin2019global}. These 4 fundamentals consisted of:

\begin{itemize}
    \item \textbf{Accountability} -- The need to stand accountable and justify one’s decisions and actions to its partners, users and others with whom the system interacts.
    \item \textbf{Governance} -- Establishment of policies, principles and/or protocols, and continuous monitoring of their proper implementations.
    \item \textbf{Explainability} -- Ensure that algorithmic decisions, as well as any data driving those decisions, can be explained to end users and other stakeholders in nontechnical terms.
    \item \textbf{Transparency} -- It must be possible to discover, trace and detect how and why a system made a particular decision or acted in a certain way, and, if a system causes harm, to discover the root cause.
\end{itemize}

It was discovered through panel discussion exercises that these categories form minimum requirement for any organization wishing to achieve eAI. In addition it was clear from within these discussions that meeting such requirements means that cross-functional considerations between roles must be taken into account from organizational levels to technical systems levels. Again, evidence continues to be collected through use-cases and further exercises to ensure these categories are exhaustive of the eAI risk landscape from societal and organisation perspectives.

\section{Proposal of a Generalized eAI Risk Assessment Methodology}
\label{sec:DRESS}

As a realisation of our results and to answer the question posed by this paper, we propose the following methodology, entitled the \textit{\textbf{Data-driven Risk Assessment Methodology for Ethical AI}} (\textbf{DRESS-eAI}). DRESS-eAI is designed to focus on the detection of pitfalls and enact the fundamentals relevant to most eAI use-cases while being structured as a process that is familiar to organizations as it is comparable to the International Organization for Standardization (ISO) standard 31000:2009 for risk management \cite{purdy2010iso}. This is an accepted standard for risk management developed by hundreds of risk management professionals over the course of four years, and has previously seen utilisation within sustainability focused methodology \cite{tiganoaia2019new}. The six process phases of the methodology are inspired directly by the IS0 31000:2009 risk management process with each phase being identified as a necessary step for systematically ensuring rigorous eAI practices of an organization. In addition, we claim that the DRESS-eAI methodology is compatible with any phase of an AI system’s life cycle, while being fully compatible with the recent Declaration of eAI issued by the AI Sustainability Center. The Declaration of eAI  can be fulfilled directly through applying DRESS-eAI to achieve fundamentals and overcome pitfalls\footnote{\url{https://aisustainability.org/the-code/}}. This declaration was issued as a response to aid organisations in preparing for the upcoming AI regulation recently proposed by the European Commission\footnote{\url{https://ec.europa.eu/info/sites/default/files/commission-white-paper-artificial-intelligence-feb2020_en.pdf}}. We envision DRESS-eAI as a formative step towards establishing the requested common normative standards for high-risk AI solutions which may pose a risk to health, safety and fundamental rights.

\begin{figure}[ht!]
    \centering
    \includegraphics[width=1\columnwidth, height=4.2in]{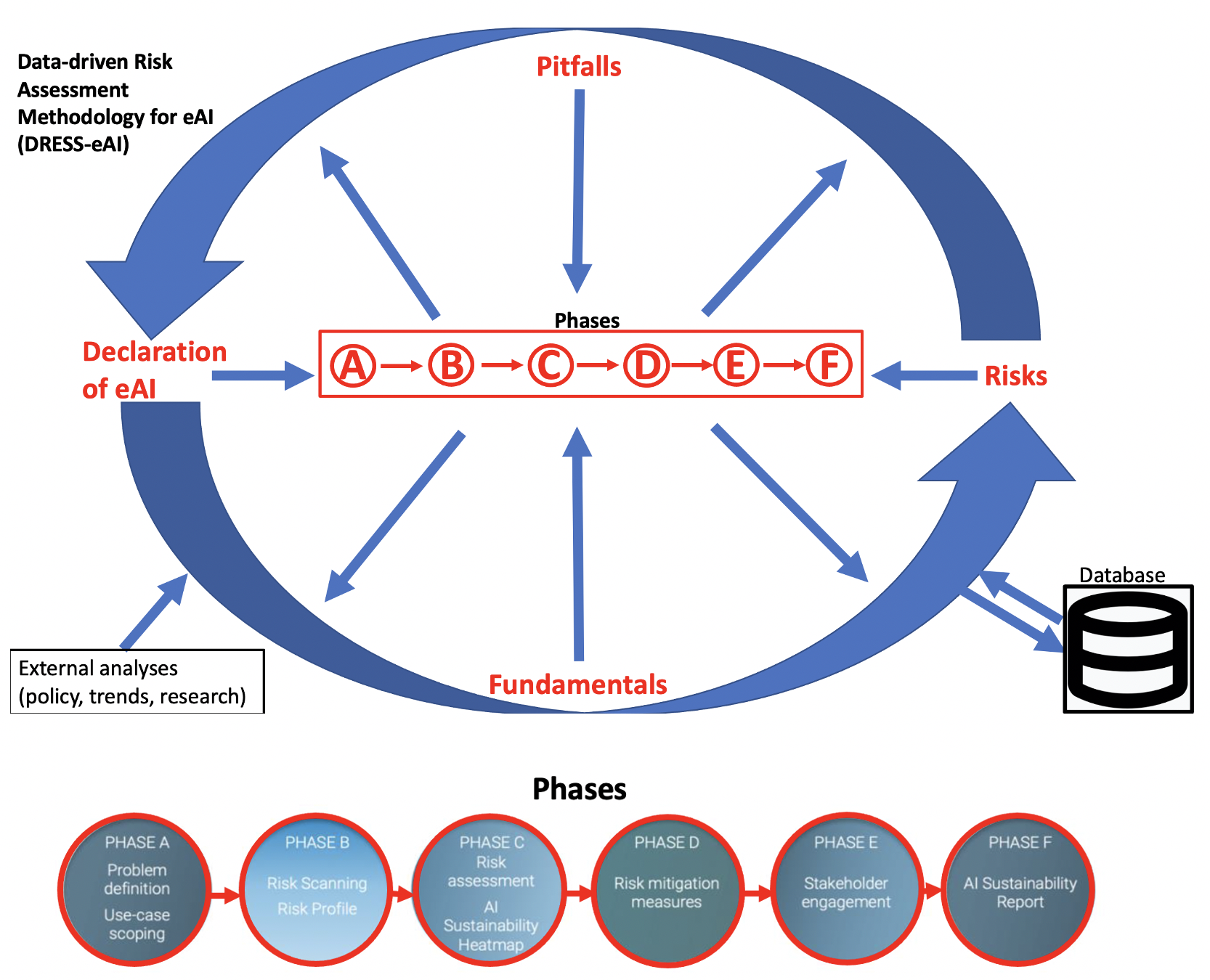}
    \vspace{-2mm}
    \caption{ An overview of DRESS-eAI. The main use-case process phases are shown in the center. Linkages show the conceptual flow between core concepts and how collected data can be used to construct and refine implementations of the phases. Also shown is how data outputs of use-cases and external analyses are recorded in a database and used for generating insights and the iterative refining of existing categories and internal processes.
    \label{fig:DRESS}}
\end{figure}

Lack of cross-functional teams tackling eAI is a thematic issue that emerged within panel discussion exercises. To accommodate for this, we advocate that all these roles/functions spanning technical, legal, risk, compliance, communications, CSR/sustainability, and HR are part of the process. Secondly, due to the specific individual risks of eAI projects, we acknowledge their relevancy in detection and mitigation as a core part of this framework. Due to the rapidly developing eAI risk landscape we also acknowledge the need for DRESS-eAI and its implementations to be evolving \textbf{data-driven projects} where generated structured data from use-cases and external analyses are recorded and used for iterative refining of existing categories and internal processes. This will permit implementations of the methodology to remain up-to-date alongside performing as a analytical tool. As such, we include a database as part of the methodology to store structured data generated from the different phases which can later be used to provide data-driven insights and refine implementations. Figure \ref{fig:DRESS}  provides a complete overview of the proposed methodology. 

We break the general process of the methodology into the following process phases for each eAI risk assessment use-case:

\begin{itemize}
    \item \textbf{Phase A:} Problem Definition/Use case scoping -- Establishing a use-case definition including summary of challenges and identifying the project team. This should achive a detailed description of the use-case, including understanding of guiding policies/codes/values, key stakeholders and technical specification. 

    \item \textbf{Phase B:} Risk Scanning/Profiling -- The key phase of the methodology. Capturing the current state of achievement towards fundamentals and vulnerability towards pitfalls with data collection form multiple organization roles. In addition, if a screening is performed from multiple perspectives this phase can provide a gap analysis which is an indication how well a specific use case is conforming to the organizational standards.
    \item \textbf{Phase C:} Risk Assessment -- Identification, evaluation and prioritization of risk scenarios. For example, workshops should be used with cross functional team to identify ethical risk scenarios and what/which stakeholders could be impacted based on the risk exposure to pitfalls and fundamentals 
    \item \textbf{Phase D:} Risk Mitigation Measures -- Identification of technical and non-technical mitigation measures and assigning ownership for actions. Identify risk mitigation measures that target the root cause of a risk scenario, or its effect. Plan for monitoring implementation of mitigating measures. 

    \item \textbf{Phase E:} Stakeholder Engagement -- Capturing Stakeholder Feedback. A necessary step to focus on those affected by the organization's activities and what is done to manage actual and potential impacts.

    \item \textbf{Phase F:} Review and maintain -- Conclusions from the completion of each phase and recommendations going forward
\end{itemize}

\subsection{Implementation of DRESS-eAI}

The above structure outlines and defines a generalized methodology for risk assessment within eAI. To explain how such a process can be enacted in the real world we explain our implementation which has been applied and refined in relation to the case studies under examination.

The chosen implementation relies on collecting structured data through self-assessment surveys. It is important to note that the chosen implementation may be prone to closed feedback loops, which can erroneously verify its own effectiveness and introduce data bias due to survey responses not reflecting true reality. As such, the implementation also collects and records qualitative feedback on the implementation directly through organizational stakeholders, permitting a deeper understanding of the implementation validity, rather than only relying on quantitative evidence acquired through repeated surveys which may possess respondent errors.

\begin{table}[ht!]

\begin{tabular}{ |p{6cm}||p{3cm}|p{3cm}|p{3cm}|  }
 \hline
 \multicolumn{4}{|c|}{\textbf{Risk Scanning Question Examples}} \\
 \hline
 \textbf{Question}  & \textbf{Fundamental} & \textbf{Pitfall} & \textbf{Organization Role}\\
 \hline
 Have you tested model results for fairness with respect to different affected groups (e.g., tested for disparate error rates)?   & Governance    & Data bias &   Technical\\
 \hline
 Have you defined what ‘human bias’ means in the context of the solution and with regards to your organizations values or policies? &   Explainability  & Bias of the creator   & CR/CSR\\
 \hline
 Are you confident in your organization's ability to detect, then shut down a malfunctioning Solution(s) in a timely manner i.e. before any harm to people or society is caused? & Governance & Misuse/Overuse &  Business owner\\
 \hline
Are the explanations that you provide about your solution easily accessible and in clear terms to external parties?     & Transparency & Misuse/Overuse &  Communications\\

 \hline
Do you have a person/function who is responsible for deciding when the algorithm(s) in the solution are mature enough/market ready?     & Accountability & Immature Data/AI &  Technical\\

 \hline
\end{tabular}
\caption{Selected examples of eAI risk scanning questions demonstrating the tagging structure of Fundamentals, Pitfall, and Organizations roles}
\label{table:1}
\end{table}

\begin{figure}[ht!]
    \centering
    \includegraphics[width=1\columnwidth, height=1.35in]{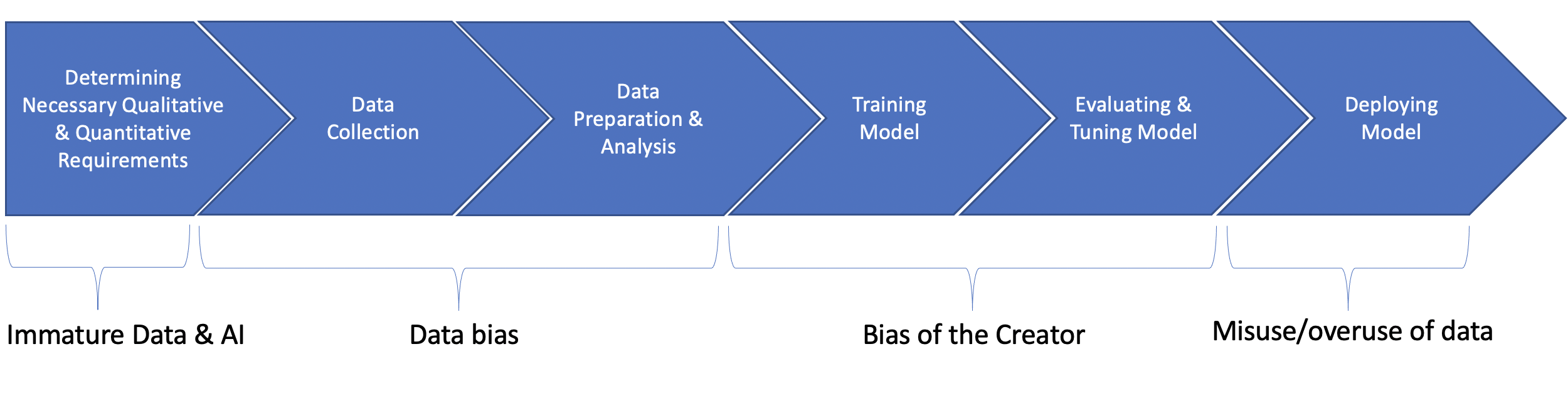}
    \vspace{-2mm}
    \caption{ Demonstrating how the pitfalls were mapped along the AI life-cycle to provide mutually exclusive categorisation of Phase B risk scan questions for the DRESS-eAI implementation. The stages shown represent universally standard steps taken by organisational teams in developing and deploying AI solutions.
    \label{fig:pitfall}}
\end{figure}

\begin{figure}[ht!]
    \centering
    \includegraphics[width=1\columnwidth, height=3.4in]{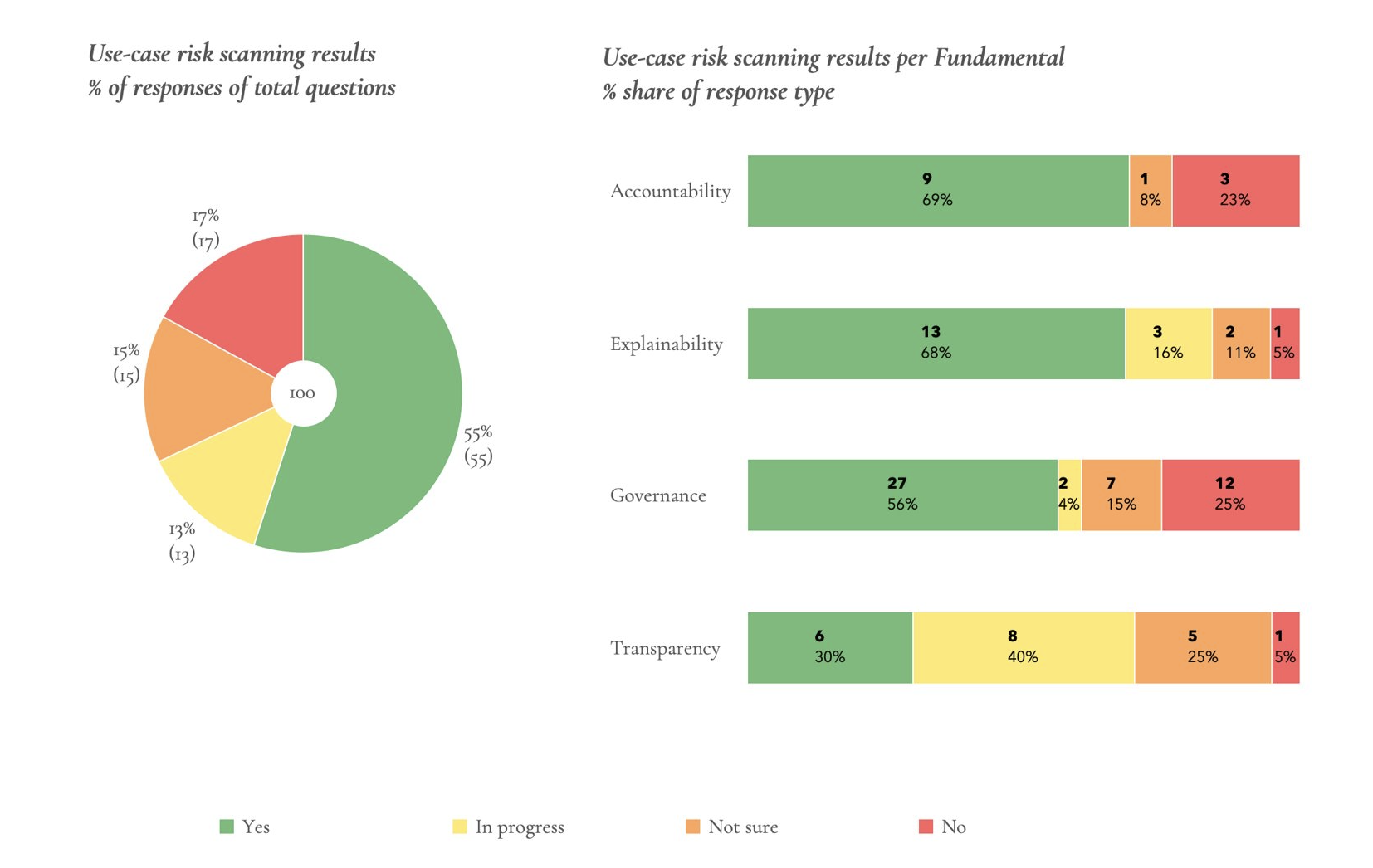}
    \vspace{-2mm}
    \caption{ Example overview of use-case output for Phase B categorized by fundamentals 
    \label{fig:example}}
\end{figure}

\begin{itemize}

    \item In \textbf{Phase A} we perform workshops for the identification detailing of an appropriate eAI use case. The outcome of this is a use-case definition including summary of challenges and detailed use-case description based on a pre-defined template. To leverage the data-driven nature of the methodology we administer three structured surveys to capture data which can later be leveraged for data-driven group-level insights between phases. Firstly, an Organizational survey to capture general questions such as the organization’s size and domain. We also administer an Organizational Maturity survey to screen for the organization’s preparedness for ethical high-risk AIs. Finally, a use case scoping survey is administered to capture a description of the AI solution that is to be assessed.

    \item In \textbf{Phase B} we cover the eAI risk landscape with an exhaustive risk scan survey of over 150 questions tagged to and equally balanced according to relevant fundamental, pitfalls and organizational role. These questions emerged as part of the same expert-based iterative process to exhaustively, and in a balanced manner, establish where an organization lies in the eAI risk landscape. This entails that each pitfall and fundamental is treated with equal priority in order to appropriately cover exposure to eAI pitfalls in the technically, legally, and societal defined risk landscape. Conceptually speaking, pitfalls may overlap with each other. However, for simplicity with our implementation, each Phase B question is tagged to a single pitfall. The decision of how to tag each question to a pitfall corresponded to which point in a AI's life cycle the question was most associated to. Figure \ref{fig:pitfall} provides an overview of how each pitfall was connected to the AI life cycle for the purposes of tagging questions. Tagging of fundamentals and roles were not associated to the AI life cycle, however, as stated, efforts were made to ensure that appropriate combinations of taggings were included to comprehensively cover the eAI risk landscape.
    
    Questions can be answered by each role with four options: 'yes', 'in-progress', 'not sure' and 'no'. We emphasize the role-based structuring of the questions to ensure the validity and comprehensiveness of answers, in addition to activating cross-functional cooperation across the organization. These roles include technical, legal, risk, compliance, communications, CSR/sustainability, business owner and HR. See Table~\ref{table:1} for examples of these questions and their tagging structure, and Figure~\ref{fig:example} for an example summary report. Importantly, all structured data from this stage is captured in our database and used to produce group-level insights surrounding the ability of this phase to exhaustively cover the risk landscape, along with how results from this phase can be linked to other phases. The output of this phase can be used to provide a gap analysis which is specified further in Section~\ref{sec:gap}

    \item In \textbf{Phase C} we identify and characterize risk scenarios guided by information attained in Phase B, constructing a traditional heatmap of risk scenarios which is prioritizes risk mitigation procedures on organizational and use-case level. Each risk scenario is tagged to a fundamental and a pitfall as well as one or more of the eight identified risks from Section~\ref{sec:research}. After a sufficient data collection period, we exploit our acquired database of risk scan surveys and risk scenarios to aid in the data-driven insight generation within and across phases. Risk scenarios are prioritized based on a qualitative analysis of likelihood and severity. Prioritized risk scenarios are characterized further, with input from additional interviews and focus meetings with the client if needed.

    \item In \textbf{Phase D} risk mitigation tools and recommendations are determined which can be technical or non-technical. We identify risk owners for the prioritized risk scenarios, either taken in the project or identified improvements needed. We provide risk mitigation from both a organisation and use-case level based on evidence acquired during evaluation. Each mitigation measure is also tagged to a fundamental and pitfall as well as one or more of the eight identified risks from Section \ref{sec:research}. After a sufficient data collection period, we leverage our database to provide data-driven recommendations and insights generation surrounding risk mitigations. Technical or non-technical risk mitigating measures are identified and implemented in broader risk management/existing processes and assigned risk owners. Examples of risk mitigation measures: updated legal documents and processes, synthetic data for avoiding bias or to preserve privacy, tailored explainability models, training, AI Ethical Principles or AI Ethical Board. Risk owners are identified within the organization and a plan is created for implementation and follow-up of actions.

    \item In \textbf{Phase E} we provide a summary of issues and recommendations on how these can be addressed and enacted through stakeholder engagements. Steps include: identifying and prioritising stakeholders to engage with; deciding what type of input is needed and whether to use existing stakeholder engagement forums /channels or targeted activities; collecting and analyzing information on key topics and problems addressed by stakeholders; stakeholder feedback captured in workshops and analyzed in a report summarizing the activities and concerns raised; potentially modified risk mitigation plan to address feedback

    \item In \textbf{Phase F} we report a summary of findings from applying the DRESS-eAI implementation. A updated risk scan of the use-case is conducted in order to track the effectiveness of risk mitigations taken over time. We also provide recommendations on how internal frameworks can be strengthened. Qualitative feedback on the implementation's true impact is acquired. 

\end{itemize}

\subsection{Gap Analysis}

\label{sec:gap}

eAI principles and commitments made by organizations are often high level, and analyses are needed to ensure a minimization of gaps between higher aspirations and what is actually happening on product and developer level \cite{mittelstadt2019principles}. Such principles ultimately have little effect on practices if they are not directly tied to structures of accountability, incentives and the ways of working in an organization. AI principles, codes and guidelines also need to be combined with monitoring of their implementation, as well as consequences if they are not met. 

\begin{figure}[ht!]
    \centering
    \includegraphics[width=0.8\columnwidth, height=3.1in]{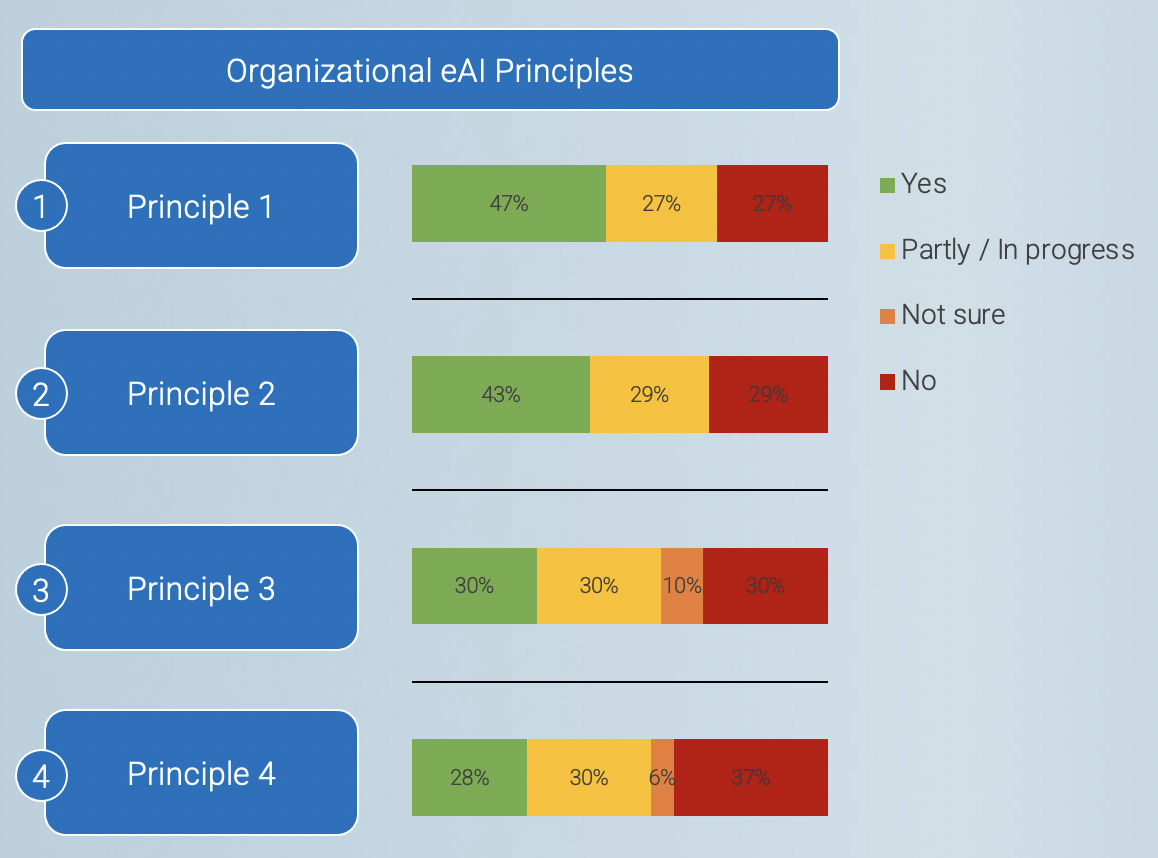}
    \vspace{-2mm}
    \caption{Example output of a gap analysis, showing how an organisation's AI principles are being achieved according to mappings to revelant DRESS-eAI's use-case risk scan questions.   
    \label{fig:gap}}
\end{figure}

The Phase B risk scanning survey output can be further used as a tool to identify possible gaps between stated ethical principles and higher aspiration and what might be happening on product or organizational level. For our DRESS-eAI implementation we also map organisational AI principles directly to risk scanning question results to facilitate the gap analysis described in Section \ref{sec:gap}. A general example of this output can be seen in Figure \ref{fig:gap}.

\section{Evaluation and Iterative Evolution of DRESS-eAI Implementation} 


As described, the DRESS-eAI risk assessment methodology has been structured to follow a data-driven iterative approach for refining implemented processes and concepts. We have implemented and tested this unique methodology for assessing AI which is compatible to typical organizational structure and usable at any point in the life-cycle of an AI-system. We propose that any implementation of our methodology should not remain a static snapshot, but a data-driven, iteratively evolving system, capturing information from each use-case for insights into the developing eAI landscape and for refining DRESS-eAI methodological implementations.

In this section we outline the application of our DRESS-eAI methodology to two real organizational case studies, reporting the effectiveness of the implementation’s ability to detect and mitigate risk, along with reporting the data-driven evolution of our implementation dictated by quantitative and qualitative evidence acquired from each case study. 

\subsection{Case Study 1: AI for assisting job hiring practices}

\subsubsection{Description} 

This use-case revolves around an AI-system being used to profile job seekers based on personal data pertaining to job hiring, education, language proficiency, as well as data about the condition and functioning of the labor market. The data used for training input was generated from various sources. For new job seekers, a self-assessment survey was answered, and personal data was generated. For job seekers already known, data was gathered from a data lake where existing data about the job seeker was stored. Job seekers are then profiled using a deep learning model on 64 features.

The output of the model was a prediction of how far from the labor market a job seeker is. Based on the outcome, job seekers are placed into three categories based on a rule-based selection. A human case worker would then be able to change the selection of which category a job seeker is placed in. However, that ability is dependent on the outcome of the profiling and automatized selection of categories. Primary early concerns were raised regarding the risk of discriminating against sensitive groups, such as foreign born women. 

\subsubsection{Outcomes}

After applying DRESS-eAI Phases A, B, and C it was identified that this use-case was exposed to several ethical and societal risks. These mainly pertained to the pitfalls ‘Bias of the Creator’ and ‘Data bias’. Weaknesses in accountability and AI governance was identified. Risk scenarios were identified and nine of them were likely to occur and could result in severe impacts on people and society. These risk scenarios were prioritized for mitigation. Mitigating measures and risk owners were identified for each risk scenario identified. The mitigating actions taken in the project were technically hardening the solution to prevent misuse, such as: updating UX interface to case workers to prevent misuse of model outputs; creating and communicating purpose statements to various stakeholders; and implementing methods for explaining model outputs and defining what needs explained and for whom (based on current and future regulation). The effect is that the solution could then be scaled. 

Furthermore, applying DRESS-eAI highlighted the need for better AI governance broadly across the organization. One key finding is a lack of ownership of an ethical AI framework internally. Following this, a cross-functional group of internal stakeholders has now been initiated as a permanent ethical AI group, with the responsibility of supporting developers of AI solutions and advancing the organizations ethical AI maturity. In addition, it is highlighted that an approach for AI fairness and explainability was needed in order to serve future AI solutions. 

\subsubsection{Input into Evolution of Implementation}

\begin{itemize}
    \item Need for separate use-case and organisational risk mitigation identified. Added distinction between risk mitigation measures that can be taken in the project vs what needs to change in the line organization. 
    \item A role was added to the survey, business owner. When implementing DRESS-eAI the need to involve representatives from business operations was identified. When an AI system is part of a larger organizational process such as in this project, many risks are associated with lack of effective collaboration and/or instructions on how to use the AI system by the business unit. Specifying a business owner role enables cross-functional collaboration in identifying potential risk exposure and taking effective mitigations. 
    \item Validation that DRESS-eAI can be applied to identify and mitigate eAI risk for a use-case in the development life-cycle phase.
\end{itemize}

\subsection{Case Study 2: AI for detecting tax fraud}

\subsubsection{Description}
This use-case began in an idea life-cycle stage, where an AI system was used to monitor and select transactions on third party market place platforms that should be reviewed for potential tax fraud. The AI system would be implemented in the third-party platforms own environment. The AI system used information about the individual and the transaction as features for classification. 

\subsubsection{Outcomes}
Applying DRESS-eAI identified which types of eAI risks can occur when using AI to detect tax fraud, lead to an increased understanding and awareness of how prepared the organization was to handle such risks. Several eAI risk aspects were highlighted including: a lack of clear organisational strategy for eAI; a lack of a systematic approach to detect and handle ethical AI risks; lack of accountability for ethical AI risks; an inability to monitor AI systems; a large exposure to pitfalls 'Data bias' and 'Bias of the creator'; A need to instate a central steering committee for overseeing eAI operations; and a need for competence development. Mitigating actions were then performed, resulting in the organisation, a year later, having an established eAI policy and plan to establish an eAI steering committee.

\subsubsection{Input into Evolution of Implementation}

\begin{itemize}
    \item Two risk scannings were required, one of the organization level and one on the standard use-case level. This lead to the inclusion of the Phase A organisational survey to aid in streamlining the implementation.
    \item Increased distinction about the difference between transparency and explainability. 
    \item Validation that DRESS-eAI can be applied to an planning stage use-case.
\end{itemize}

\section{Leveraging Group Level Results – Refinement of data collection tool, Data Strategy and Insights}

In this section we examine the the data-driven aspect of the DRESS-eAI from two perspectives.  Firstly, we demonstrate how repeated application of use-cases have guided the refinement of our implementation. Secondly, we outline the general data strategy of our implementation to provide a better understanding of how group-level data can and should be leveraged to refine implementations and provide eAI insights both within and between DRESS-eAI phases. 

\subsection{Refinement of survey tool through use-case insights}

Any implementation of DRESS-eAI will demand the refinement of process tools to better accommodate for organisational needs and the developing eAI landscape. For our implementation, we recorded common feedback attained during use-cases in our database and made refinements based on group-level evidence. To help demonstrate this process, we present representative examples of how repeated use-cases of our implementation resulted in the refinements of our data collection process. A complete overview of these examples can be viewed in Table \ref{table:3}.

\textit{In the first example}, we examine a question pertaining to the data bias pitfall and governance fundamental. The question pertains to the needs of building a solution on the same data distributions in which it will be deployed to help ensure it is not simply well fit to a training data set and then under-performs on unfamiliar data examples in the real world. The original question was reported as not being concrete in it's intentionally in this regard. Since a technical role was intended to answer such a question, more detailed terminology about data sets and statistical distributions was included. In general, more exact terms for technical roles questions were added across questions.

\textit{In the second example}, we see an addition question was added to help better capture the product owner's input on whether they oversee the compatibility of the solution to their organisation's values. This was part of the general refinement of the the risk scan to have the organisational product owner more involved in the risk scanning. This inclusion was noted as being crucial as product owners tended to understand the intended use and value of the solution more than other roles. To better identify potential vulnerabilities, more questions were added to have the product owner role as a larger part within the risk scan process, specifically asking them more organisationally related questions surrounding the governance and accountability pitfalls. 

\textit{In the third example}, it was reported that the original question could lead to incorrect responses due to ambiguities. The intention of the original question was to establish from the product owner whether or not they possess a general open data collection strategy; a lack of which could lead to insufficient data in terms of quality. However, the original wording of the question could lead respondents to understand this as a question simply related to communicating human biases for selecting data. This update represents an example in which ambiguities in the questions were removed. 

\textit{In the fourth example}, it was noted that the question was both ambiguous and interpreted  incorrectly by respondents. The intention of the question was to highlight a data bias vulnerability due to having automated data bias strategy. This could lead to data bias due to a lack of human oversight. The original wording of the question did not make this intention clear as thus lead to incorrect responses. Such a question is representative of similar questions that needed to be rephrased.

\textit{In the fifth example}, we highlight a general case in which redundant questions needed to be removed. For the questions shown  we noted from feedback that the general organisation question was sufficiently answered with a similar question and thus could be removed.

\begin{table}[H]

\begin{tabular}{ |p{6cm}||p{3cm}|p{3cm}|p{3cm}|  }
 \hline
 \multicolumn{4}{|c|}{\textbf{Risk Scanning Question Example 1 Version 1}} \\
 \hline
 \textbf{Question}  & \textbf{Fundamental} & \textbf{Pitfall} & \textbf{Organization Role}\\
 \hline
     Is the distribution of demographic groups in your dataset representative of the reality you are trying to reflect? & Governance    & Data bias &   TECH\\

 \hline
\end{tabular}

\begin{tabular}{ |p{6cm}||p{3cm}|p{3cm}|p{3cm}|  }
 \hline
 \multicolumn{4}{|c|}{\textbf{Updated Risk Scanning Question Example 1}} \\
 \hline
 \textbf{Question}  & \textbf{Fundamental} & \textbf{Pitfall} & \textbf{Organization Role}\\
 \hline
     Is the distribution of demographic groups in your dataset representative of the  distribution present in the population(s) where your solution is/are deployed    & Governance    & Data bias &   TECH\\

 \hline
\end{tabular}
\end{table}

\begin{table}[H]

\begin{tabular}{ |p{6cm}||p{3cm}|p{3cm}|p{3cm}|  }
 \hline
 \multicolumn{4}{|c|}{\textbf{Risk Scanning Question Example 2 Version 1}} \\
 \hline
 \textbf{Question}  & \textbf{Fundamental} & \textbf{Pitfall} & \textbf{Organization Role}\\
 \hline
    - (Question needed to be added)     & -    & - &   -\\

 \hline
\end{tabular}

\begin{tabular}{ |p{6cm}||p{3cm}|p{3cm}|p{3cm}|  }
 \hline
 \multicolumn{4}{|c|}{\textbf{Updated Risk Scanning Question Example 2}} \\
 \hline
 \textbf{Question}  & \textbf{Fundamental} & \textbf{Pitfall} & \textbf{Organization Role}\\
 \hline
    Are you as product owner involed in the design, development, auditing etc. of the solution to ensure that the solution conforms to your organizationial values     & Accountability    & Misuse/Overuse &   Product owner\\

 \hline
\end{tabular}
\end{table}

\begin{table}[H]

\begin{tabular}{ |p{6cm}||p{3cm}|p{3cm}|p{3cm}|  }
 \hline
 \multicolumn{4}{|c|}{\textbf{Risk Scanning Question Example 3 Version 1}} \\
 \hline
 \textbf{Question}  & \textbf{Fundamental} & \textbf{Pitfall} & \textbf{Organization Role}\\
 \hline
  Do you communicate to relevant stakeholders about on what biases and values your data was selected and processed?    & Transparency    & Data bias &   Product owner\\

 \hline
\end{tabular}

\begin{tabular}{ |p{6cm}||p{3cm}|p{3cm}|p{3cm}|  }
 \hline
 \multicolumn{4}{|c|}{\textbf{Updated Risk Scanning Question Example 3}} \\
 \hline
 \textbf{Question}  & \textbf{Fundamental} & \textbf{Pitfall} & \textbf{Organization Role}\\
 \hline
   Do you communicate to relevant stakeholders about on what grounds the data was selected and processed?     & Transparency    & Data bias &   Product owner\\

 \hline
\end{tabular}

\end{table}

\begin{table}[H]

\begin{tabular}{ |p{6cm}||p{3cm}|p{3cm}|p{3cm}|  }
 \hline
 \multicolumn{4}{|c|}{\textbf{Risk Scanning Question Example 4 Version 1}} \\
 \hline
 \textbf{Question}  & \textbf{Fundamental} & \textbf{Pitfall} & \textbf{Organization Role}\\
 \hline
     Do you have an automated process for data validation?     & Governance    & Data bias &   TECH\\
 \hline
\end{tabular}

\begin{tabular}{ |p{6cm}||p{3cm}|p{3cm}|p{3cm}|  }
 \hline
 \multicolumn{4}{|c|}{\textbf{Updated Risk Scanning Question Example 4}} \\
 \hline
 \textbf{Question}  & \textbf{Fundamental} & \textbf{Pitfall} & \textbf{Organization Role}\\
 \hline
    Do you have an automated process/approach with human oversight for data validation?     & Governance    & Data bias &   TECH\\
 \hline
\end{tabular}
\end{table}

\begin{table}[H]

\begin{tabular}{ |p{6cm}||p{3cm}|p{3cm}|p{3cm}|  }
 \hline
 \multicolumn{4}{|c|}{\textbf{Risk Scanning Question Example 5 Version 1}} \\
 \hline
 \textbf{Question}  & \textbf{Fundamental} & \textbf{Pitfall} & \textbf{Organization Role}\\
 \hline
Do you have a diversity policy and procedures to ensure diversity in your organization?     & Governance & Bias of the creator &  HR\\

 \hline
Do you have processes/approaches in place to ensure that there is diversity within your pool of designers and managers involved in the creation of the solution in terms of gender, culture, age, etc.?      & Governance & Bias of the creator &  HR\\

 \hline
\end{tabular}

\begin{tabular}{ |p{6cm}||p{3cm}|p{3cm}|p{3cm}|  }
 \hline
 \multicolumn{4}{|c|}{\textbf{Updated Risk Scanning Question Example 5}} \\
 \hline
 \textbf{Question}  & \textbf{Fundamental} & \textbf{Pitfall} & \textbf{Organization Role}\\
 \hline
- (Question removed due to overlap)     & Governance & Bias of the creator &  HR\\

 \hline
Do you have processes/approaches in place to ensure that there is diversity within your pool of designers and managers involved in the creation of the solution in terms of gender, culture, age, etc.?      & Governance & Bias of the creator &  HR\\

 \hline
\end{tabular}
\caption{Selected examples of eAI risk scanning questions showing the iterative evolution using case study evidence}
\label{table:3}
\end{table}

\subsection{Data Strategy and Insights}

Importantly, we wish to highlight the relevancy for this methodology to be data-driven by possessing a underlying database capable of storing structured information from each use-case to attain group-level insights. Utilising this data-driven backbone of the methodology permits refinements in terms of efficiency, effectiveness, and deployability across a variety of contexts. We also model DRESS-eAI as process that is concurrently refined though complementary external input such as new regulations, trends, and research. This part of the data-driven backbone which permits continuous adaption of implementations to the rapidly changing eAI landscape.

In practical terms, insights attained from group-level data are exploited to improve each implementation in the following manner:  

\begin{itemize}
    \item Providing summary reports on the general effectiveness of an implementation, and the state of the eAI landscape.
    \item Permitting the benchmarking of eAI organisational status on per-sector and cross-sector levels.
    \item Acquiring greater contextual information with less time burden on clients through personalized questions   
    \item Identifying deficiencies with existing surveys or tools.
    \item Developing internal and client dashboards and PR reports 
\end{itemize}

With the DRESS-eAI structure, data insights can be attained though the independent analysis of each implementation phase. Of equal importance is the potential to understand how data from each phase is connected. Within the implementation applied for this study we utilise a general approach for mapping phases together which can provide a structure for attaining informative results though the means of statistical analysis and AI modeling. For our implementation, we build data relations between various phases through common attributes for each output data table.  More specifically, we achieve such relations by tagging all questions in Phase B, risks scenarios in Phase C, and mitigation measures in Phase D with attributes of their respective pitfalls and fundamentals. Phase C risk scenarios and Phase D risk mitigations are also tagged to the eight risk categories identified and defined as part of this study. See Section \ref{sec:research} for an overview of the eight risks.

\section{Conclusion and Future Directions}
\label{sec:conclusion}

In this paper we have outlined and motivated the problem of developing a well-vetted and real-world applicable approach to ethical AI risk assessment. We report the findings of our systematic multidisciplinary research approach to building definitions and establishing requirements needed for such a methodology. Importantly our approach to involve cross-sector experts has highlighted a need for a methodology that incorporates cross-functional considerations that builds on familiar organisational processes. Leveraging this evidence, we then propose a novel methodology named DRESS-eAI. Furthermore we fully describe our implementation and report the effectiveness and evolution of our implementation by describing several case studies and group-level insights. 

As ongoing work we are actively employing the implementation of DRESS-eAI with organisations; continuously acquiring evidence to understand how our implementation of the methodology can be further refined. Such evidence will permit additional group level analyses, afforded by the data-driven backbone of DRESS-eAI, which will be leveraged for providing data-driven insights, while refining risk assessment tools and gap analyses.

\appendix

 \bibliographystyle{elsarticle-num} 
 \bibliography{Manuscript}





\end{document}